# New Algorithmic Approaches for Computing Optimal Network Paths with Several Types of QoS Constraints

Mugurel Ionuţ Andreica, Nicolae Ţăpuş
Computer Science and Engineering Department
Politehnica University of Bucharest
Bucharest, Romania
e-mail: {mugurel.andreica, nicolae.tapus}@cs.pub.ro

*Abstract*—The problem of efficiently delivering data within networks is very important nowadays, especially in the context of the large volumes of data which are being produced each year and of the increased data access needs of the users. Efficient data delivery strategies must satisfy several types of Quality of Service (QoS) constraints which are imposed by the data consumers. One possibility of achieving this goal (particularly in the case of in-order data transfers) is to choose a satisfactory network delivery path. In this paper we present novel algorithmic approaches for computing optimal network paths which satisfy several types of (QoS) constraints.

*Keywords-optimal constrained network path; QoS*

## I. INTRODUCTION

Efficient delivery of data is an important issue nowadays, particularly in the context in which large amounts of data are produced, accessed and consumed every year. Moreover, data consumers impose stringent QoS constraints on the data delivery mechanism in order to consider it of a good enough quality. In the case of in-order data delivery, one possibility of satisfying the constraints is to choose an adequate network delivery path. In this paper we consider multiple constrained network path optimization problems, for which we present novel algorithmic solutions. In Section II we consider an approximate bicriteria optimal path problem. In Section III we present sensitivity analysis algorithms for several optimal path problems (and more). In Section IV we present novel solutions for several Hamiltonian and constrained color-alternating paths and cycles in graphs. In Section V we discuss related work and in Section VI we conclude.

## II. APPROXIMATE BICRITERIA OPTIMAL PATH

We consider a directed graph $G$ with $n$ vertices and $m$ edges. Each directed edge $e$ from $u(e)$ to $v(e)$ has two non-negative integer weights $w_1(e)$ and $w_2(e)$. We want to find a path from a source vertex $s$ to a destination vertex $t$ which minimizes the sum of the $w_1$ weights of the edges on the path, under the constraint that the sum of the $w_2$ edges of the path is at most $B$ (at least $B$). A standard but inefficient solution to this problem consists of constructing a graph $G'$ of pairs $(i,j)$, where $i$ is a vertex of $G$ and $j$ is an integer value between $0$ and $Bmax$=the sum of all the $w_2$ weights of the edges of $G$. For every edge $e$ from $u$ to $v$ we add $Bmax-w_2(e)+1$ edges, from a vertex $(u,j)$ to the vertex $(u, j+w_2(e))$ ($0 \leq j \leq Bmax-w_2(e)$); each such edge has a weight equal to $w_1(e)$. We now only need to compute the shortest paths (using the weights of the edges of $G'$) from $(s,0)$ to all the other vertices. The answer to our problem is the minimum length of the shortest path from $(s,0)$ to some $(t,j)$, with $j \leq B$ (or $j \geq B$). A much more efficient approach, albeit a heuristic one, is the following. We assign to every edge $e$ of $G$ a cost function $c(e,x)=w_1(e)+x \cdot w_2(e)$, where $x$ is a parameter. We compute the shortest path from $s$ to $t$ using the newly assigned cost functions of the edges, for $x=0$ and we compute $w_2sum(0)$=the sum of the $w_2$ weights of the edges on the path. If $w_2sum(0) \leq B$ (and we are in the *"at most B"* case) or $w_2sum(0) \geq B$ (and we are in the *"at least B"* case), then the shortest path at $x=0$ is the one we are looking for (because it minimizes the sum of the $w_1$ weights disregarding the $w_2$ weights; however, the sum of the $w_2$ weights of the edges on the path satisfies the constraints). Let's assume now that $w_2sum(0)>B$ and we are looking for the path from $s$ to $t$ minimizing the sum of the $w_1$ weights with the constraint that the sum of the $w_2$ weights is at most $B$. We notice that as the parameter $x$ increases from $0$ to $+\infty$, the total sum of the $w_2$ weights on the shortest path from $s$ to $t$ (using the cost functions $c(*,x)$) decreases (although not strictly). In fact, we have that $w_2sum(x') \geq w_2sum(x'')$ if $x' \leq x''$. At $x=+\infty$, the influence of the $w_1$ weights is negligible. Thus, at $x=+\infty$, we can consider $c(e,x=+\infty)=w_2(e)$ for every edge $e$. Let $w_2sum(+\infty)$ be the sum of the $w_2$ weights of the shortest path from $s$ to $t$ when $x=+\infty$. If $w_2sum(+\infty)>B$ then we have no solution. Otherwise, we will binary search the smallest value of $x$ (between $0$ and $+\infty$) for which $w_2sum(x) \leq B$. The sum of the $w_1$ weights on the edges of the shortest path from $s$ to $t$ for the value of $x$ we found in the binary search is the minimum possible sum of the $w_1$ weights on a path from $s$ to $t$ which satisfies the constraints and which can be found by this algorithm. Note that $w_2sum(x)$ is computed as follows. We compute the shortest path from $s$ to $t$ using the costs $c(e,x)$ on the edges. Then, we find the edges composing the actual path and we let $w_2sum(x)$ be the sum of the $w_2$ weights of these edges. The time complexity is $O((m+n \cdot log(n)) \cdot log(x))$ for arbitrary directed graphs and $O((m+n) \cdot log(x))$ for directed acyclic graphs.

In the second case, when we want the sum of the $w_2$ weights to be at least $B$ and $w_2sum(0)<B$, we notice that as we decrease $x$ from $0$ towards $-\infty$, the value of $w_2sum(x)$ increases, i.e. we have $w_2sum(x') \geq w_2sum(x'')$ if $x' \leq x''$.

Unlike the previous case, we now have to handle negative edge weights. We will first need to find the minimum value of $xmin$ for which the graph $G$ with the edge costs $c(e,xmin)$ does not contain a negative cost cycle (we can first check if $G$ contains a negative cost cycle for $x=-\infty$; note that $c(e,x=-\infty)=-w_2(e)$). If it doesn't, then we set $xmin=-\infty$. Otherwise, we start with $xm=-1$ and, as long as the graph does not have negative cost cycles for $x=xm$, we multiply $xm$ by any value $c>1$ (i.e. we set $xm=c\cdot xm$; e.g. $c=2$). Then, we binary search $xmin$ on the interval $[xm,0]$. After this, we compute $w_2sum(xmin)$. If $w_2sum(xmin)<B$ then we stop. Otherwise, we will binary search the largest value of $x$ in the interval $[xmin,0]$ for which $w_2sum(x)\geq B$. The sum of the $w_1$ weights on the shortest path from $s$ to $t$ in the graph with edge costs $c(e,x)$ where $x$ is the value found by the binary search is the minimum sum of the $w_1$ weights of a path from $s$ to $t$ which satisfies the constraints (and which can be found by this algorithm). The time complexity in this case is $O(m\cdot n\cdot log(x))$ for arbitrary directed graphs (because of the negative edge costs, we need to use algorithms like Bellman-Ford-Moore) or $O((m+n)\cdot log(x))$ for acyclic directed graphs.

Notice that in the case of directed acyclic graphs we can also compute the path from $s$ to $t$ with maximum sum of the $w_1$ weights, such that the sum of the $w_2$ weights is at most $B$ (at least $B$) with the same $O((m+n)\cdot log(x))$ time complexity. We introduce the same edge costs and we compute $w_2sum(0)$ like before (except that we now run a longest path algorithm instead of a shortest path algorithm). If $w_2sum(0)$ satisfies the constraints, then the initial path we found is the optimal solution. Otherwise, if $w_2sum(0)>B$ (for the case *"at most B"*), we have the property that $w_2sum(x')\leq w_2sum(x'')$ for $x'\leq x''$. Thus, we will find a value $xmin<0$ such that $w_2sum(xmin)\leq B$ (the same way we found the value $xmin$ previously). Then, we will binary search the largest value of $x$ in the interval $[xmin,0]$ such that $w_2sum(x)\leq B$. The longest path corresponding to this value of $x$ is a near-optimal path for our problem. If, initially, we have $w_2sum(0)<B$ (for the case *"at least B"*), we will first find a value $xmax>0$ such that $w_2sum(xmax)\geq B$ (we start with $xmax=1$ and we multiply it by $c>1$ until the condition is met). Then, we binary search the smallest value of $x$ in the interval $[0,xmax]$ for which $w_2sum(x)\geq B$. The longest path corresponding to this value of $x$ is a (near) optimal path for our original problem.

An optimal path (with respect to the aggregation function *aggf*) from every vertex $u$ to $t$ in a directed acyclic graph in which every edge $e$ and vertex $v$ have a weight $we(e)$ and $wv(v)$ can be computed recursively, using memoization. We initialize the values *computed(u)=false* for every vertex $u$. We have *popt(t)=wv(t)* and *computed(t)=true*. Then, we consider every vertex $u$ and we call *Compute(u)* which takes the following steps: *(1)* if *computed(u)=false* then: { *(1.1) popt(u)=aggf(wv(u), opt{aggf(we(e), Compute(v)) | e is an edge directed from u to v});* *(1.2) computed(u)=true;* } *(2)* return *popt(u)*. *opt* can be, for instance, *max* or *min*.

III.    SENSITIVITY ANALYSIS ALGORITHMS

*A.  Classifying Graph Edges and Vertices*

We consider a directed graph $G$, in which every directed edge $(u,v)$ (from vertex $u$ to vertex $v$) has a weight $w(u,v)$. We would like to classify the vertices and edges of this graph into *3* categories, relative to the shortest path (i.e. the path of minimum total weight) between two given vertices $s$ and $t$: *1)* belonging to every shortest path from $s$ to $t$ ; *2)* belonging to at least one shortest path from $s$ to $t$ ; *3)* belonging to no shortest path from $s$ to $t$. We will start by computing the shortest path tree rooted at $s$ (i.e. the shortest paths from $s$ to every other vertex of the graph). We can achieve this in $O(m+n\cdot log(n))$ time, where $m$ is the number of edges of the graph and $n$ is the number of vertices. Let $ds(u)$ be the distance from $s$ to $u$ in the graph (if $u$ is not reachable from $s$, then $ds(u)=+\infty$). If $ds(t)=+\infty$ then $t$ is not reachable from $s$ and all the edges and vertices belong to category *3* (as there is no path from $s$ to $t$).

If $ds(t)<+\infty$ then we compute the shortest path distances from $t$ to every vertex of the graph, considering the transposed graph (i.e. the graph in which the direction of each edge is reversed). Let $dt(u)$ be the distance from $t$ to $u$ in the transposed graph. We will mark the vertices $u$ for which $ds(u)+dt(u)=ds(t)(=dt(s))$; the other vertices will be left unmarked. Note that $s$ and $t$ are also marked (as $ds(s)=dt(t)=0$ and $ds(t)=dt(s)$). Then, we construct the graph $G'$ containing all the marked vertices and all the edges between them (an edge $(u,v)$ is included in $G'$ only if both $u$ and $v$ were marked). Every edge that was not included in $G'$ and every vertex that was not marked belong to category *3*. After this, we will ignore the direction of the edges of $G'$, thus turning $G'$ into an undirected graph $G''$. In $G''$ we will compute the bridges and cut vertices, in $O(m+n)$ time. Every directed edge $(u,v)$ from G' corresponding to a bridge in $G''$ belongs to category *1*; every vertex from $G'$ corresponding to a cut vertex in $G''$ also belongs to category *1*. All the other edges and vertices from $G'$ belong to category *2*.

When the edge weights are all *1* (or all equal, in which case we can replace them by *1*), we can improve the time complexity to $O(m+n)$. The distances from $s$ to every vertex of $G$ and from $t$ to every vertex of the transpose of $G$ can be computed with a simple BFS traversal. Afterwards, we construct $G'$ like before (again, all the edges and vertices outside of $G'$ belong to category *3*). After constructing $G'$, we do not need to construct $G''$ anymore and compute its edges and cut vertices. Instead, we will initialize an array *cnt* (with indices from *0* to *n-1*) to *0*. Then, for every vertex $u$ of $G'$ we increase *cnt(ds(u))* by *1*. *cnt(d)* will be equal to the number of vertices $u$ of $G'$ such that $ds(u)=d$. A vertex $u$ of $G'$ belongs to category *1* only if *cnt(ds(u))=1* (otherwise, if *cnt(ds(u))>1*, then $u$ belongs to category *2*). A directed edge $(u,v)$ of $G'$ belongs to category *1* only if both $u$ and $v$ belong to category *1*; otherwise, $(u,v)$ belongs to category *2*.

The algorithms described in this section can also be used for undirected graphs. If we have an undirected graph $UG$, we will construct a directed graph $G$ with the same vertices as $UG$, by adding to $G$ two directed edges $(u,v)$ and $(v,u)$ for every undirected edge $(u,v)$ of $UG$; both edges will have the same weight $w(u,v)$. Then, we run one of the algorithms described above. A vertex of $UG$ belongs to the same category as the corresponding vertex of $G$. An edge $(u,v)$ of $UG$ belongs to category *1* (*2*) if at least one of the edges $(u,v)$ or $(v,u)$ belongs to category *1* (*2*). Note that if one of the edges $(u,v)$ and $(v,u)$ belongs to category *1* or *2*, then the other edge belongs to category *3*. An edge $(u,v)$ of *UG* belongs to category *3* if both edges $(u,v)$ and $(v,u)$ of *G* belong to category *3*.

*B. Classifying Items Relative to Knapsack Solutions*

We have *n* items. Each item *i* ($1 \leq i \leq n$) has a weight $w(i) \geq 0$. We want to find a subset of items whose sum of weights is exactly *S* (the knapsack problem). We would like to classify the *n* items into *3* categories: *1)* belonging to every knapsack solution; *2)* belonging to at least one knapsack solution; *3)* belonging to no solution. We will assume that we can afford $O(n \cdot S)$ time and memory. We will compute $ok_1(i,j)=true$, if we can obtain the sum *j* using some of the items *1, ..., i*, or *false*, otherwise ($0 \leq i \leq n$; $0 \leq j \leq S$). Moreover, we also compute $cnt_1(i,j)$=the number of solutions for obtaining the sum *j* from the items *1, ..., i* (bounded from above at any value $Q \geq 2$, even $Q=+\infty$; $0 \leq i \leq n$; $0 \leq j \leq S$). We have $ok_1(0,0)=true$, $cnt_1(0,0)=1$, $ok_1(0,1 \leq j \leq S)=false$ and $cnt_1(0,1 \leq j \leq S)=0$. For $1 \leq i \leq n$ (in this order) we have: $ok_1(i,0 \leq j < min\{w(i),S+1\})=ok_1(i-1,j)$, $cnt_1(i,0 \leq j < min\{w(i),S+1\})=cnt_1(i-1,j)$), $ok_1(i,w(i) \leq j \leq S)=ok_1(i-1,j)$ or $ok_1(i-1, j-w(i))$ (the first case considers that item *i* is not used for obtaining the sum *j*, while the second case considers that item *i* is used for obtaining the sum *j*) and $cnt_1(i,w(i) \leq j \leq S)=min\{Q, cnt_1(i-1,j)+cnt_1(i-1,j-w(i))\}$. Then, in a similar manner, we compute $ok_2(i,j)=true$, if we can obtain the sum *j* using some of the items *i, i+1, ..., n*, or *false*, otherwise, and $cnt_2(i,j)$ ($1 \leq i \leq n+1$; $0 \leq j \leq S$). We have $ok_2(n+1,0)=true$, $cnt_2(n+1,0)=1$, $ok_2(n+1,1 \leq j \leq S)=false$ and $cnt_2(n+1,1 \leq j \leq S)=0$. For $1 \leq i \leq n$ (in decreasing order) we have: $ok_2(i,0 \leq j < min\{w(i), S+1\})=ok_2(i+1,j)$, $cnt_2(i,0 \leq j < min\{w(i), S+1\})=cnt_2(i+1,j)$, $ok_2(i,w(i) \leq j \leq S)=ok_2(i+1,j)$ or $ok_2(i+1,j-w(i))$ and $cnt_2(i,w(i) \leq j \leq S)=min\{Q, cnt_2(i+1,j)+cnt_2(i+1,j-w(i))\}$. With these tables, we can decide for each item *i* in $O(S)$ time to which category it belongs. We will compute $c(i)$=the sum of the values $cnt_1(i-1,j) \cdot cnt_2(i+1,S-j)$ ($0 \leq j \leq S$), and $d(i)$=the logical *OR* of the values ($ok_1(i-1,j)$ and $ok_2(i+1,S-j-w(i))$) ($0 \leq j \leq S-w(i)$; $d(i)=false$ if $w(i)>S$). If $d(i)=false$ then item *i* belongs to category *3*; otherwise, if $c(i)>0$ then item *i* belongs to category *2* (there is at least one solution without item *i*) and if $c(i)=0$ then item *i* belongs to category *1* (there is no solution without item *i*, meaning that item *i* belongs to every knapsack solution).

Note how this solution is much better than the trivial solution which would remove every item *i* one at a time, would recompute the knapsack problem's dynamic programming tables every time (obtaining an $O(n^2 \cdot S)$ time complexity), and would compute $a(i)=true$ (*false*) if the sum $S-w(i)$ can (cannot) be obtained and $b(i)=true$ (*false*) if the sum *S* can (cannot) be obtained. If $a(i)=false$ then item *i* is in category *3*; if $a(i)=true$ and $b(i)=true$ ($b(i)=false$) then item *i* is in category *2* (*1*).

We will now also assign a cost *cost(i)* to every item *i* and we want to find a subset of items whose sum of weights is equal to *S* and whose sum of costs is minimum. As before, we want to classify the items into the same *3* categories, relative to an optimal solution for this problem: *1)* belonging to every optimal solution; *2)* belonging to at least one optimal solution; *3)* belonging to no optimal solution. We will compute $cmin_1(i,j)$ ($cmin_2(i,j)$) = the minimum cost of a subset of the items *1, ..., i* (*i, ..., n*) such that their sum of weights is *j* ($0 \leq j \leq S$); we also compute $cnt_1(i,j)$ ($cnt_2(i,j)$) = the number of subsets, bounded from above by any value $Q \geq 2$ (even $Q=+\infty$), of the items *1, ..., i* (*i, ..., n*) whose sum of weights is *j* and whose sum of costs is $cmin_1(i,j)$ ($cmin_2(i,j)$) ($0 \leq j \leq S$). We have $cmin_1(0,0)=cmin_2(n+1,0)=0$, $cnt_1(0,0)=cnt_2(n+1,0)=1$, $cmin_1(0,1 \leq j \leq S)=cmin_2(n+1, 1 \leq j \leq S)=+\infty$, and $cnt_1(0,1 \leq j \leq S)=cnt_2(n+1,1 \leq j \leq S)=0$.

For $1 \leq i \leq n$ (in increasing order) we have: $cmin_1(i, 0 \leq j < min\{w(i), S+1\}) = cmin_1(i-1,j)$, $cnt_1(i,0 \leq j < min\{w(i), S+1\})=cnt_1(i-1, j)$, $cmin_1(i,w(i) \leq j \leq S)=min\{cmin_1(i-1,j), cmin_1(i-1,j-w(i))+cost(i)\}$, and $cnt_1(i,w(i) \leq j \leq S)=min\{Q$, the sum of the values $cnt_1(i-1,j)$ and (*if* ($cmin_1(i-1, j-w(i))+cost(i)=cmin_1(i,j))$ *then* $cnt_1(i-1, j-w(i))$ *else* $0$)*}*. Similarly, for $1 \leq i \leq n$ (in decreasing order) we have: $cmin_2(i, 0 \leq j < min\{w(i), S+1\}) = cmin_2(i+1,j)$, $cnt_2(i, 0 \leq j < min\{w(i), S+1\})=cnt_2(i+1, j)$, $cmin_2(i, w(i) \leq j \leq S) = min\{cmin_2(i+1,j), cmin_2(i+1, j-w(i))+cost(i)\}$, and $cnt_2(i, w(i) \leq j \leq S)=min\{Q$, the sum of the values $cnt_2(i+1,j)$ and (*if* ($cmin_2(i+1, j-w(i))+cost(i)=cmin_2(i,j))$ *then* $cnt_2(i+1, j-w(i))$ *else* $0$)*}*. With these tables, we will decide for each item *i* into which category it belongs. If $cmin_1(n,S)=+\infty$ then all the items belong to category *3*. Otherwise, for each item *i*, we will compute the following values: $c(i)$=the sum of the values $cnt_1(i-1,j) \cdot cnt_2(i+1,S-j)$ (where $0 \leq j \leq S$ and $cmin_1(i-1,j)+cmin_2(i+1,S-j)=cmin_1(n,S))$, and $d(i)=min\{cmin_1(i-1,j)+cost(i)+cmin_2(i+1,S-j-w(i)) \mid 0 \leq j \leq S-w(i)\}$ (if $w(i)>S$ then $d(i)=+\infty$). If $d(i)>cmin_1(n,S)$ then the item *i* belongs to category *3*. Otherwise (if $d(i)=cmin_1(n,S))$ then: if $c(i)>0$ then item *i* belongs to category *2*, otherwise (if $c(i)=0$) then item *i* belongs to category *1*.

IV. HAMILTONIAN AND COLOR-ALTERNATING PATHS AND CYCLES

*A. Hamiltonian Path in a Tournament Graph*

We consider a tournament graph *G* (a directed graph in which there is exactly one directed edge between any two vertices *i* and *j*: either the edge *i->j*, or *j->i*) with *n* vertices. We want to find a Hamiltonian path in this graph. The graph is not given explicitly. Its structure can be discovered by asking questions: *Ask(u,v)* ($u \neq v$) returns *1*, if the edge *u->v* exists in the graph, or *-1*, if the edge *v->u* exists in *G*.

A tournament graphs always contains at least one Hamiltonian path. We can construct such a path incrementally. We will start with a path of length *1*, consisting of the vertex *1*. Then, we will traverse the other vertices *i=2,...,n*. When we reach the vertex *i*, we will already have a path *v(1), ..., v(i-1)* composed of the vertices *1, ..., i-1* (in some order), such that we have the directed edges *v(j)->v(j+1)* ($1 \leq j \leq i-2$). If the edge *i->v(1)* exists in *G*, then we can add the vertex *i* before *v(1)* and obtain a path with *i* vertices. If, instead, the edge *v(i-1)->i* exists in *G*, then we can add *i* after *v(i-1)* and obtain a path with *i* vertices. If none of the edges *i->v(1)* and *v(i-1)->i* exist in

$G$, then we will traverse the vertices $v(j)$ on the path found so far ($j=2,...,i-1$), until we find the first vertex for which the edge $i\rightarrow v(j)$ exists in $G$. We will insert the vertex $i$ between $v(j-1)$ and $v(j)$, obtaining a new path: $v(1), ..., v(j-1), i, v(j), ..., v(i-1)$. The time complexity of this approach is $O(n^2)$. Another $O(n^2)$ solution is the following. We consider the vertices in the order $p(1)=1, ..., p(i)=i, ..., p(N)=N$. Then, we will repeatedly traverse these vertices. When we find two adjacent vertices $p(i)$ and $p(i+1)$ such that the edge between them is oriented from $p(i+1)$ to $p(i)$, then we swap the two vertices in the $p$ ordering. When no more swaps can be performed, we stop. This algorithm is very similar to bubble-sort. The previous algorithm was very similar to the insertion sort method. The $O(n^2)$ solutions presented earlier can be improved to $O(n \cdot log(n))$. We will start by improving the first presented solution, as follows. Let's assume that we arrived at vertex $i$ and we currently have a path $v(1), ..., v(i-1)$ (containing the vertices $1, ..., i-1$). We will find in $O(log(n))$ time the place where the vertex $i$ will be inserted in the path. If we have the edge $i\rightarrow v(1)$, then we add the vertex $i$ before $v(1)$; otherwise, if we have the edge $v(i-1)\rightarrow i$ then we add vertex $i$ after $v(i-1)$. Otherwise, we know that we have the edges $v(1)\rightarrow i$ and $i\rightarrow v(i-1)$. We will use the binary search technique and, at every step, we will maintain an interval $[a,b]$, such that we have the edges $v(a)\rightarrow i$ and $i\rightarrow v(b)$. Initially, $a=1$ and $b=i-1$. Within the binary search, we will select $c=(a+b)\ div\ 2$. If we have the edge $v(c)\rightarrow i$, then we maintain further the interval $[c,b]$; otherwise, we have the edge $i\rightarrow v(c)$ and we will maintain further the interval $[a,c]$. The binary search ends when $b=a+1$. Then, we insert vertex $i$ between $v(a)$ and $v(a+1)$ in the path.

Other solutions with $O(n \cdot log(n))$ time complexities are based on sorting algorithms like merge-sort or quick-sort. Basically, we consider that every vertex $i$ of the graph has an associated value $val(i)$. When running the sorting algorithms, we will try to sort the vertices in "increasing" order of their values. When we need to compare the values of two vertices $i$ and $j$, we make the following decision:
- if we have the edge $i\rightarrow j$ then $val(i)<val(j)$
- if we have the edge $j\rightarrow i$ then $val(j)<val(i)$

Although the tournament graph may contain cycles, the ordering produced by the sorting algorithms using the decision process described above sorts the graph vertices in an order $v(1), ..., v(n)$, such that we have the edges $v(i)\rightarrow v(i+1)$ ($1\leq i\leq n-1$).

### B. Constrained Path/Cycle in a Small Graph

We consider a directed graph with $n$ vertices, where $n$ is not too large. Every directed edge $e$ (directed from $u(e)$ to $v(e)$) has a cost $cm(e)\geq 0$ and every vertex $u$ has a cost $cn(u)\geq 0$. We want to find a path/cycle containing exactly $Q$ vertices for which the aggregate $agg$ of the edge and vertex costs is minimum. For $Q=1$ we simply choose the vertex $u$ with the minimum value $cn(u)$. For $Q\geq 2$ we will compute the values $D_{min}(i, S)$=the minimum cost of a path ending at the vertex $i$ and passing through every vertex from the set $S$ ($S$ is a subset of $\{1,...,n\}$ and contains the vertex $i$) and through no other vertex; we will also compute $C_{min}(i,j,S)$=the minimum cost of a path whose two end-vertices are $i$ and $j$ (the path is directed from $i$ to $j$) and passes through every vertex of the set $S$ ($S$ is a subset of $\{1,...,n\}$ and contains the vertices $i$ and $j$) and through no other vertex. We will consider the subsets $S$ in increasing order of their cardinality $|S|$ (from $1$ to $Q$). We have $D_{min}(i, \{i\})=cn(i)$ and $C_{min}(i,i,\{i\})=cn(i)$. For $|S|\geq 2$, we will proceed as follows. $D_{min}(i,S)$ (with $i \in S$) is equal to $cn(i)\ agg\ min\{D_{min}(j, S\setminus\{i\})+cm(e)\ |\ e$ is an edge directed from a vertex $j$ to the vertex $i$, $1\leq j\leq n$, $j\neq i$, $j\in S\setminus\{i\}\}$. $C_{min}(i, j, S)$ is equal to $min\{U(i,j,S), V(i,j,S)\}$, where: $U(i, j, S)=cn(j)\ agg\ min\{C_{min}(i, k, S\setminus\{j\})\ agg\ cm(e)\ |\ e$ is an edge directed from the vertex $k$ to the vertex $j$, $1\leq k\leq n$, $k\neq j$, $k\in (S\setminus\{j\})\}$ and $V(i, j, S)=cn(i)\ agg\ min\{Cmin(k, j, S\setminus\{i\})\ agg\ cm(e)|e$ is an edge directed from the vertex $i$ to the vertex $k$, $1\leq k\leq n$, $k\neq i$, $k\in (S\setminus\{i\})\}$ (actually, $V(i,j,S)$ may always be considered $+\infty$). The answer will be $min\{D_{min}(i,S)|1\leq i\leq N, |S|=Q\}$ (for the path case), respectively $min\{C_{min}(i,j,S)\ agg\ cm(e)\ |\ e$ is an edge directed from $j$ to $i$, $1\leq i,j\leq N$, $i\neq j$, $|S|=Q\}$. The time complexity of the algorithm is $O(n^2 \cdot T(n,Q))$ for the path case and $O(n^3 \cdot T(n,Q))$ for the cycle case, where $T(n,Q)= C(n,0)+C(n,1)+...+C(n,Q)$ ($C(i,j)$=combinations of $i$ elements taken as $j$). When $Q=n$, $T(n,n)=2^n$. Note that the algorithm can be generalized as follows: only some subsets $S$ of vertices are allowed to form the path/cycle; in this case, we simply iterate when computing the answer over the valid subsets $S$ only.

### C. Hamiltonian Path in the Cube of a Graph

The cube graph $G^3$ of an undirected graph $G$ is the graph in which two vertices $x$ and $y$ are directly connected by an edge if their distance in $G$ is at most $3$. In order to find a Hamiltonian path in $G^3$, we will first find a spanning tree $T$ of $G$ and then we will focus on finding a Hamiltonian path in $T^3$. We will choose an arbitrary root vertex $r$ for $T$ and we will perform a DFS traversal of $T$ starting from $r$. We consider the level of each node: $level(r)=1$ and $level(i\neq r)= level(parent(i))+1$. While traversing the tree, we will construct the Hamiltonian path $HP$ as follows. We start with an empty path $HP$. Then, during the traversal, when we first enter a vertex $i$ and $level(i)$ is odd, we add the vertex $i$ at the end of $HP$ (before traversing any of the other vertices in vertex $i$'s subtree). After finishing traversing the entire subtree of a vertex $i$, if $level(i)$ is even, then we add $i$ at the end of $HP$. It is easy to notice that the distance in $T$ between any two consecutive nodes in $HP$ is at most $3$: if they have the same level parity, the distance between them is $2$ (they are either two sons of the same vertex or one of them is the grand-parent of the other); the distance between two consecutive vertices $x$ and $y$ from $HP$ can also be $3$, by considering the following scenario – $level(x)$ is even and we exit the subtree of the vertex $x$, we return to vertex $x$'s grand-parent $z$ and then we enter the vertex $y$ which is one of $z$'s sons; of course, the distance may also be $1$.

## D. Optimal Color Alternating Path

We consider a graph with $n$ vertices and $m$ directed edges. Each edge $e$ is directed from $u(e)$ to $v(e)$, has a cost $cost(e) \geq 0$ and a color $col(e)$ (e.g. label). The colors are numbered from $1$ to $C$. Every vertex $u$ also has a cost $cn(u)$. We want to find a path from the vertex $s$ to the vertex $t$ whose aggregate cost is minimum (using the aggregation function $agg$) and such that any two consecutive edges on the path have different colors. The first solution consists of constructing a color-expanded graph, in which every node is a pair $(i,k)$, meaning that we reached node $i$ of the initial graph and the last used edge had color $k$. Normally, $1 \leq k \leq C$, but we will also consider the case $k=0$ for the vertex $i=s$. For every directed edge $e$ of the original graph we add an edge from $(u(e),k)$ to $(v(e),col(e))$ with the cost $cost(e)+cn(v(e))$ ($1 \leq k \leq C$; $k$ may also be $0$ when $u(e)=s$; $k \neq col(e)$). We will now compute the tree of minimum aggregate paths starting from $(s,0)$ in the color-expanded graph (considering costs only on the graph's edges). The minimum cost of a color-alternating path from $s$ to $t$ is $cn(s)$ agg the minimum of the costs of reaching a vertex $(t,k)$ ($0 \leq k \leq C$). The color-expanded graph has $O(n \cdot C)$ vertices and $O(m \cdot C)$ edges. Another possibility is the following. We will compute the minimum cost of a path from $s$ to every other vertex $i$ of the graph (let this value be $C_{min}(i)$). We will also maintain $Col_{min}(i)$=the color of the last edge on the optimal path from $s$ to $i$. We also compute $C_{min,2}(i)$=the minimum cost of a path from $s$ to $i$ such that the color of its last edge is different from $Col_{min}(i)$; let the color of the last edge of the path denoted by $C_{min,2}(i)$ be $Col_{min,2}(i)$. If the graph is directed and acyclic, then we will first compute a topological sort. Then, we traverse the vertices $i$ in the order of this sort. For a vertex $i$ we will consider as candidate *(Cost, Color)* pairs the following values: $(C_{min}(u(e))$ agg $cost(e)$ agg $cn(i), col(e))$ (where $v(e)=i$ and $col(e) \neq Col_{min}(u(e))$) and $(C_{min,2}(u(e))$ agg $cost(e)$ agg $cn(i), col(e))$ (where $v(e)=i$ and $col(e) \neq Col_{min,2}(u(e))$). If $i=s$ then $C_{min}(s)=cn(s)$, $Col_{min}(s)=0$, $C_{min,2}(s)=+\infty$ and $Col_{min,2}(s)=0$. For $i \neq s$ we initialize $C_{min}(i)=C_{min,2}(i)=+\infty$ and $Col_{min}(i)=Col_{min,2}(i)=0$. For each candidate pair *(Cost, Color)*: *(1)* if $Cost < C_{min}(i)$ then: *(1.1)* if $Color \neq Col_{min}(i)$ then we set $(C_{min,2}(i), Col_{min,2}(i))=(C_{min}(i), Col_{min}(i))$; *(1.2)* $(C_{min}(i), Col_{min}(i))=(Cost, Color)$; *(2)* if $Cost \geq C_{min}(i)$ and $Cost < C_{min,2}(i)$ and $Color \neq Col_{min}(i)$ then we set $(C_{min,2}(i), Col_{min,2}(i))=(Cost, Color)$. If the graph is not acyclic, then we could use the Bellman-Ford-Moore shortest path algorithm. We initialize the computed values like before. Then, we insert the vertex $s$ in a queue $Q$. While $Q$ is not empty, we extract the vertex $i$ from the front of $Q$ and we process it. We consider all the edges $e$ with $u(e)=i$. The pairs $(C_{min}(i)$ agg $cost(e)$ agg $cn(v(e)), col(e))$ (if $Col_{min}(i) \neq col(e)$) and $(C_{min,2}(i)$ agg $cost(e)$ agg $cn(v(e)), col(e))$ (if $Col_{min,2}(i) \neq col(e)$) are candidate pairs for the vertex $v(e)$. For each candidate pair we proceed as before. If at least one of the values $C_{min}(v(e))$ or $C_{min,2}(v(e))$ changes as a result of these actions, then we (re)insert the vertex $v(e)$ into $Q$ (if it is not already there).

## E. Color Alternating Euler Cycle

We consider an undirected graph with $n$ vertices and $m$ edges. Every edge $e$ (between $u(e)$ and $v(e)$) has a color $col(e)$. We want to find an Euler cycle in this graph such that any two edges which are consecutive on the cycle have different colors. At first, we will consider every vertex $x$ of the graph, together with all the edges $e$ adjacent to it. Let's assume that the edge identifiers are unique, between $1$ and $m$. The edges adjacent to the vertex $x$ will be considered in an arbitrary order: $mu(x,j)$, with $1 \leq j \leq deg(x)$ (where $deg(x)$ denotes the degree of the vertex $x$ and $mu(x,j)$ is an edge identifier). Moreover, we will maintain a hash table $Hx(x)$ associated to every vertex $x$, where we will introduce pairs of the form $(key=mu(x,j),\ value=j)$. The only condition required for having a color alternating Euler cycle is for the graph to be connected, the degree of every vertex must be even, and the number of edges with the same color adjacent to any vertex $x$ should be at most $deg(x)/2$. Next, we will assign to every edge $mu(x,j)$ another edge $mu(x,pair(x,j))$, such that $mu(x,j)$ and $mu(x,pair(x,j))$ have different colors (and also $mu(x,pair(x,\ pair(x,j)))=mu(x,j)$). In order to achieve this, let's consider the following problem. We have $2 \cdot P$ objects of different colors, such that there are at most $P$ objects of the same color. We want to split the objects into pairs such that the two objects of a pair should have different colors. Let $C$ be the total number of distinct colors assigned to the $2 \cdot P$ objects. We will renumber the colors from $1$ to $C$. We traverse the objects and maintain a hash table $H$ (initially empty) and a counter $C$ (initially $0$). For every color $C'$ of an object, we first look the key $C'$ up in $H$; if we find it there, then let $col$ be the value associated to the key $C'$ in $H$ – we will replace the color $C'$ of the object by the color $col$. If $C'$ is not in $H$, then we increment $C$ by $1$, we add to $H$ the pair *(key=C', value=C)* and then we replace the color $C'$ of the object by the color $C$. After this, we will construct an array $nob(col)$=the number of objects having the color $col$ (we initialize all the values of the array to $0$; then, we traverse all the $2 \cdot P$ objects and, for every color $col$ of an object, we increment $nob(col)$ by $1$). Then, we will sort the $C$ colors $col$ in non-increasing order, according to their $nob(col)$ values (we will use count sort in order to sort all the $C$ colors in $O(C)$ time, or any other sorting algorithm, for $O(C \cdot log(C))$ time). Let this order be $col_1, ..., col_C$. Then we sort the objects in the order $Ob(1), ..., Ob(2 \cdot P)$, such that the objects with the color $col_i$ are located on consecutive positions, starting from the position $nob(col_1)+...+nob(col_{i-1})+1$, and ending at the position $nob(col_1)+...+nob(col_i)$. We can perform this sort very easily, in $O(P)$ time, by constructing some lists $L(col)$=the list with all the objects having the color $col$ ($1 \leq col \leq C$), which we will then concatenate in the order $L(col_1), ..., L(col_C)$. The $P$ pairs we will construct will contain the objects $(Ob(i), Ob(P+i))$ ($1 \leq i \leq P$). It is obvious that two objects $Ob(i)$ and $Ob(P+i)$ cannot have the same color. Using the solution to this problem and considering as colored objects the edges adjacent to every vertex $x$, we can assign to every edge

$mu(x,j)$ another edge $mu(x,pair(x,j))$ (we will also have $mu(x,pair(x,pair(x,j)))=mu(x,j)$).

The second stage of the algorithm consists of incrementally constructing the Euler cycle, as follows. For each vertex $x$ we will mark all the edges $mu(x,j)$ adjacent to $x$ which were added to the cycle so far (initially, all the edges are unmarked) and we will maintain an index $muidx(x)$, representing the fact that all the edges $mu(x,1), …, mu(x,muidx(x))$ have already been added to the cycle (initially, $muidx(*)=0$). We will maintain a linked list with the graph vertices and edges in the order in which they occur on the cycle. Between every two list elements corresponding to the vertices of the cycle we will have a list element corresponding to the edge connecting the two vertices (for the case where we can have multiple edges between the same pair of vertices). Initially, we will only introduce in the list an element corresponding to the vertex $1$ (we consider the vertices numbered from $1$ to $n$). Then, we will traverse the list. Let's assume that we reached an occurrence of a vertex $x$ within the list. While $muidx(x)<deg(x)$ we will perform the following actions: *(Step 1)* if the edge $mu(x, muidx(x)+1)$ has already been marked as belonging to the cycle, then we simply increment $muidx(x)$ by $1$; *(Step 2)* if the edge $mu(x,muidx(x)+1)$ has not been marked as belonging to the cycle, then we will construct a cycle starting from this edge. Let's assume that this edge connects the vertex $x$ to the vertex $y$. At first, we mark the edge $mu(x, muidx(x)+1)$. Then, we set $nodc=y$ and $mprev=mu(x, muidx(x)+1)$ and we will construct the cycle as follows. While $nodc \neq x$ or the value associated to the edge $mprev$ in $Hx(x)$ is not $pair(x,muidx(x)+1)$ then: *(a)* $midxprev$=the index (value) associated to the key $mprev$ in $Hx(nodc)$; *(b)* $midxnext=pair(nodc, midxprev)$; *(c)* $mnext=mu(nodc, midxnext)$; *(d)* we mark the edge $mnext$ as belonging to the cycle; *(e)* the edge $mnext$ connects the vertex $nodc$ to another vertex $z$; *(f)* we set $nodc=z$ and $mprev=mnext$. At the end of the cycle construction, we are back at vertex $x$, the first edge of the cycle is $mu(x,muidx(x)+1)$ and the last edge of the cycle is $mu(x,pair(x,muidx(x)+1))$. We will construct this cycle as a linked list, too (which is initialized with the elements corresponding to the vertex $x$, the edge $mu(x,muidx(x)+1)$, and then the vertex $y$; after every execution of the step *(f)* we add to the list an element corresponding to the edge $mprev$ and then an element corresponding to the vertex $nodc$). We will try to replace the occurrence of the vertex $x$ from the main cycle (the occurrence from which we started to construct the new cycle), by the newly constructed cycle. If $x$ is the first vertex from the list corresponding to the main cycle, then we can realize this immediately. Otherwise, let $m_1$ and $m_2$ be the edges preceding and succeeding $x$ in the list of the main cycle. If $m_1$'s color is different than the color of the edge $mu(x,muidx(x)+1)$ and $m_2$'s color is different from that of the edge $mu(x,pair(x,muidx(x)+1))$, then we can replace the occurrence $x$ by the list corresponding to the new cycle right away (we only need to adjust the linked list pointers in order to insert the new cycle within the list of the main cycle instead of the former occurrence of the vertex $x$). If the condition does not hold, then we can insert the new cycle in the main cycle list in reverse order: instead of the current occurrence of the vertex $x$ in the main cycle list, we will insert the newly constructed cycle, but traversed in the other direction than the one in which it was constructed (i.e. in reverse order) – thus, in the main cycle, $m_1$ will be followed by the edge $mu(x,pair(x,muidx(x)+1))$ and the edge $mu(x,muidx(x)+1)$ will be followed by the edge $m_2$. After performing these actions we will consider that the current occurrence of the vertex $x$ is the one corresponding to the first element of the newly constructed cycle. After this, we finished the execution of the *Step 2* and we can go to the next iteration of the *"while muidx(x)<deg(x)"* loop.

## V. RELATED WORK

Efficient algorithms for a parametric shortest path problem which is similar to the one we presented in Section II were given in [1]. Sensitivity analysis algorithms for edges and vertices relative to matchings (in bipartite graphs) and minimum spanning trees were given in [2]. Algorithms for finding Hamiltonian paths in tournament graphs have been known for quite some time [3], but our approach describes multiple solutions and presents more details than in other publications. Hamiltonian paths and cycles in the cube of a tree were previously studied, for instance, in [4]. The novelty of our approach is the algorithm itself. The color alternating Euler cycle has also been studied before in [5], but they use there different methods for matching pairs of edges adjacent to the same vertex.

## VI. CONCLUSIONS

In this paper we presented novel algorithmic techniques for several constrained optimal path problems in networks. We considered both bicriteria and single criterion optimization problems. We also developed efficient algorithms for the sensitivity analysis of network links and vertices. Many of the presented algorithms can be applied in real-life scenarios, in which the network parameters are not too dynamic.